\documentclass[prl,twocolumn, superscriptaddress]{revtex4-2}

\usepackage{graphicx} 
\usepackage{amssymb} 
\usepackage[usenames]{color}
\usepackage{amsmath}
\usepackage{xspace}
\usepackage{longtable}
\usepackage{soul}
\usepackage{epstopdf}
\usepackage{array, makecell}
\usepackage{tabularx}
\usepackage{physics}

\usepackage[colorlinks = true,linkcolor = blue,urlcolor  = blue,citecolor = blue,anchorcolor = blue]{hyperref}

\newcommand{\Mntriangle}{Mn(Me$_6$tren)$X$\xspace}
\newcommand{\Efield}{$E$-field\xspace}
\newcommand{\df}{$\delta f_E$\xspace}

\begin{document}

\title{Chemical tuning of quantum spin-electric coupling in molecular nanomagnets} 

\author{Mikhail V. Vaganov}
\thanks{These three authors contributed equally}
\affiliation{CAESR, Department of Physics, University of Oxford, The Clarendon Laboratory, Parks Road, Oxford OX1 3PU, UK}

\author{Nicolas Suaud}
\thanks{These three authors contributed equally}
\affiliation{Laboratoire de Chimie et Physique Quantiques (LCPQ), Universit\'{e} de Toulouse, CNRS, 118 route de Narbonne, F-31062 Toulouse, France}

\author{Fran\c{c}ois Lambert}
\thanks{These three authors contributed equally}
\affiliation{Institut de Chimie Mol\'{e}culaire et des Mat\'{e}riaux d'Orsay, CNRS, Universit\'{e} Paris-Saclay, 17, avenue des Sciences, 91400 Orsay, France}

\author{Benjamin Cahier}
\affiliation{Laboratoire de Chimie et Physique Quantiques (LCPQ), Universit\'{e} de Toulouse, CNRS, 118 route de Narbonne, F-31062 Toulouse, France}

\author{Christian Herrero}
\affiliation{Institut de Chimie Mol\'{e}culaire et des Mat\'{e}riaux d'Orsay, CNRS, Universit\'{e} Paris-Saclay, 17, avenue des Sciences, 91400 Orsay, France}

\author{R\`{e}gis Guillot}
\affiliation{Institut de Chimie Mol\'{e}culaire et des Mat\'{e}riaux d'Orsay, CNRS, Universit\'{e} Paris-Saclay, 17, avenue des Sciences, 91400 Orsay, France}

\author{Anne-Laure Barra}
\affiliation{Laboratoire National des Champs Magnétiques Intenses, UPR CNRS 3228, Univ. Grenoble Alpes, 25, avenue des Martyrs, B.P. 166, 38042 Grenoble Cedex 9, France}

\author{Nathalie Guih\'{e}ry}
\email{nathalie.guihery@irsamc.ups-tlse.fr}
\affiliation{Laboratoire de Chimie et Physique Quantiques (LCPQ), Universit\'{e} de Toulouse, CNRS, 118 route de Narbonne, F-31062 Toulouse, France}

\author{Talal Mallah}
\email{talal.mallah@universite-paris-saclay.fr}
\affiliation{Institut de Chimie Mol\'{e}culaire et des Mat\'{e}riaux d'Orsay, CNRS, Universit\'{e} Paris-Saclay, 17, avenue des Sciences, 91400 Orsay, France}

\author{Arzhang Ardavan}
\email{arzhang.ardavan@physics.ox.ac.uk}
\affiliation{CAESR, Department of Physics, University of Oxford, The Clarendon Laboratory, Parks Road, Oxford OX1 3PU, UK}

\author{Junjie Liu}
\email{junjie.liu@physics.ox.ac.uk}
\affiliation{CAESR, Department of Physics, University of Oxford, The Clarendon Laboratory, Parks Road, Oxford OX1 3PU, UK}
\affiliation{School of Physical and Chemical Sciences, Queen Mary University of London, London E1 4NS, UK}

\begin{abstract}

\textbf{Controlling quantum spins using electric rather than magnetic fields promises significant architectural advantages for developing quantum technologies. In this context, spins in molecular nanomagnets offer tunability of spin-electric couplings (SEC) by rational chemical design. Here we demonstrate systematic control of  SECs in a family of Mn(II)-containing molecules via chemical engineering. The trigonal bipyramidal (tbp) molecular structure with $C_3$ symmetry leads to a significant molecular electric dipole moment that is directly connected to its magnetic anisotropy. The interplay between these two features gives rise to significant experimentally observed SECs, which can be rationalised by wavefunction theoretical calculations. Our findings guide strategies for the development of electrically controllable molecular spin qubits for quantum technologies.}

\end{abstract}
\maketitle
The possibility of electrical spin control offers significant architectural advantages for classical or quantum spintronics because, compared to magnetic fields, electric fields can be efficiently routed and confined in complex nanoscale circuits, thereby reducing energy consumption and facilitating logic operations on spins~\cite{Kane1998,Manipatruni2019,Long2020,Ramesh2021,Yang2023}. Research into interactions between electric fields and spin degrees of freedom in various quantum systems have attracted interest both theoretically and experimentally~\cite{Trif2008,Thiele2014,Baumann2015,Asaad2019,Pradines2022,Zhang2022,Wang2023}. A strong SEC is critical both for efficient electrical quantum spin control and for engineering coherent spin-electric interfaces allowing exchange of quantum information between distinct spin qubits~\cite{Mi2017,Yu2023}. 

Among the candidates for spin qubits, molecular nanomagnets offer particular advantages: coordination chemistry allows rich tunability of molecular quantum spin structures while also providing routes to large-scale integration via supramolecular approaches~\cite{Atzori2019,Gaita-Arino2019,Wasielewski2020}. One approach to enhancing SECs in molecular nanomagnets \cite{Robert2019,Kintzel2021,Liu2020,Liu2021} is to exploit strong spin-orbit coupling (SOC) by employing heavy-metals, e.g. rare earth atoms, as the spin carrier. For example, a Ho(III)--containing molecular nanomagnet, in which a small structural distortion from strict tetragonal symmetry leads to an \Efield sensitive spin transition, demonstrates a SEC that is sufficiently strong to enable selective spin control using modest electric fields of $10^5$ V/m~\cite{Liu2021}. Although providing important insights, the Ho(III) example also highlights a limitation of this approach: such molecules typically possess a giant zero-field splitting (ZFS), leading to inconveniently large transition energies between spin states. Furthermore, the origin of the SEC in this system is an accidental symmetry-breaking, rather than the result of rational chemical design. Therefore, we identify the challenge of engineering molecular nanomagnets with spin transitions that are both in an accessible energy range and sensitive to electric fields.

The $S = 5/2$ spin associated with a Mn(II) ion is a simple quantum system with potential for quantum information processing (QIP).  As a free ion, it has a half-filled $3d$ shell with the electron ground state of $S = 5/2$ and $L = 0$. In molecular or crystalline environments, the weak SOC leads to small magnetic anisotropies, reduced spin-lattice relaxation and impressive spin relaxation times. It also removes one of the key ingredients for enhancing SEC, at first sight compromising the scope for efficient \Efield spin control~\cite{Fittipaldi2018,Fang2022,Liu2024}. Indeed, so far, a sizable SEC in Mn(II) has only been observed when doped into ferro- or piezo-electric hosts~\cite{George2013} offering little scope for tuning the spin properties. 

In this work we exploit the chemical control over the coordination sphere available in a family of molecular nanomagnets containing Mn(II) to engineer SECs. Our strategy is to design a molecular geometry, namely [Mn(me$_6$tren)$X$]$Y_2$ (where me$_6$tren is Tris[2-(dimethylamino)ethyl]amine, $X$ = Cl, $Y$ = ClO$_4$ (\textbf{1}), $X$ = Br, $Y$ = PF$_6$ (\textbf{2}) and $X$ = I, $Y$ = I (\textbf{3})), which, by virtue of its substantial in-built electric dipole, exhibits a significant \Efield-induced deformation that is also coupled to the molecular spin anisotropy. This approach yields SECs comparable with those only observed so far for lanthanide-based molecules with strong SOC.

Our rational design approach allows us to tune the SEC by varying the coordination environment of the spin centre systematically. Wavefunction-based \textit{ab initio} calculations suggest that the molecular ZFS originates from competing contributions from excited electronic states with distinct symmetries. The large electric dipole along the threefold symmetry axis allows strong modifications of these contributions by an electric field, leading to significant SECs. The strongest effect is observed for \textbf{3} where both the electric dipole moment and the \Efield-induced deformation are the largest.

\begin{section}{Results and Discussion}

\begin{subsection}{The \Mntriangle compounds}

\begin{figure*}[t]
\centering
\includegraphics[width=2\columnwidth]{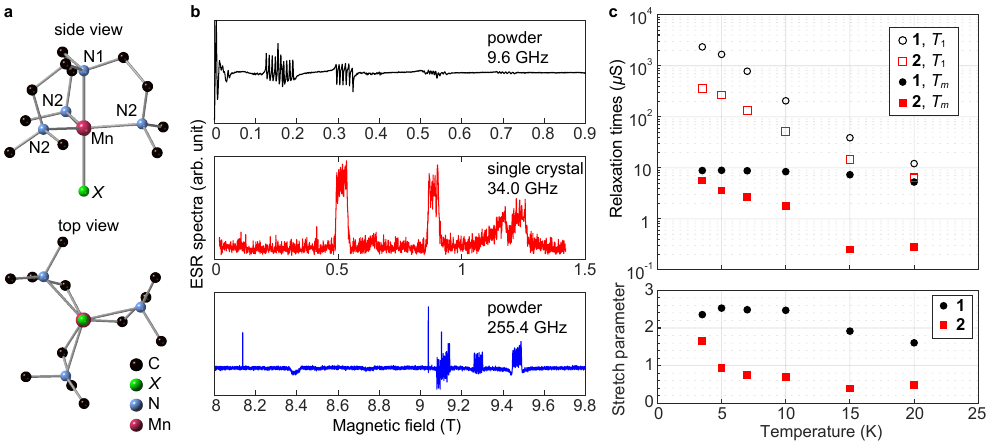}
\caption{\textbf{(a)} Ball-and-stick representation of the [Mn(me$_6$tren)$X$] molecules. H atoms are omitted for clarity. \textbf{(b)} Representative low-temperature ESR spectra for \textbf{1} recorded with different sample forms at different frequencies.  The single crystal spectrum (middle) was recorded at Q-band using an echo-detected field sweep (EDFS) whereas the ESR experiments for the powder sample (top and bottom) were conducted using the continuous-wave method. \textbf{(c)} Low-temperature relaxation times for MnCl and MnBr molecules measured on the $-5/2 \leftrightarrow -3/2$ and $+3/2 \leftrightarrow +5/2$ transitions, respectively. (\textbf{c, upper panel}) The spin-lattice relaxation time, $T_1$, and quantum phase memory time, $T_\mathrm{m}$, for MnCl and MnBr as a function of temperature.  $T_1$ is desribed by a single exponential decay over the experimental temperature range.  In contrast, $T_\mathrm{m}$ follows a stretched exponential,  whose  stretch parameter varies with temperature as shown in \textbf{(c, lower panel)}.}
\label{fig_relaxation}
\end{figure*}

The crystallographic structure of the Cl derivative (\textbf{1}) is the same as that reported for the Ni(II) and Co(II) counterparts~\cite{Ruamps2013,Ruamps2014,Rubin-Osanz2021}. None of the reported Br derivatives with any metal ion crystallize in a trigonal space group. We, therefore, prepared a new compound with Br in the axial position and PF$^{6-}$ as counter anion that turned out to crystallize in a trigonal space group (see below). For the I derivative, despite trying several counter anions, they all crystallize in a cubic space group. We therefore prepared the Mn complex based on the reported Zn(II) one that crystallizes in a cubic space group \cite{Naveen2021}. Mn(II) is pentacoordinate surrounded by one axial (N1) and three equatorial (N2) nitrogen atoms belonging to the tetradentate me$_6$tren ligand, and one halogen ($X$). Its coordination sphere has a trigonal bipyramidal (tbp) geometry of $C_3$ point group symmetry, with the threefold axis is along the N1-Mn-$X$ direction (Fig.~\ref{fig_relaxation}\textbf{a}). The Mn-N1, Mn-N2 bond lengths and the N1MnN2 angles differ by less than one percent for the three complexes. The main difference is the Mn-$X$ distance:  2.346~\AA, 2.503~\AA and 2.713~\AA for $X$ = Cl, Br and I, respectively. \textbf{1} and \textbf{2} crystallise in the $R_3C$ and the $R_3$-2 trigonal space group, with the $C_3$ molecular axis along the crystal $c$ axis and all the N1-Mn-$X$ bonds aligned. \textbf{3} crystallises in the cubic $P2_13$ space group with the $C_3$ molecular axes along the cubic unit cell diagonal. We used the isostructural diamagnetic Zn(II)-containing compounds to provide a diamagnetic host crystal with dilute Mn(II)-complex impurities (see SI).

We characterised the magnetic properties using electron spin resonance (ESR) at three frequencies; representative data for \textbf{1} are shown in Fig.~\ref{fig_relaxation}\textbf{b} (see SI for more data). The results can be described by an electron spin $S = 5/2$ and a nuclear spin $I = 5/2$ under the Hamiltonian
\begin{equation}
\label{H_Mn}
\hat{H} = D\hat{S}_z^2 + \mu_Bg\mathbf{B}_0\cdot\hat{\mathbf{S}} + A\hat{\mathbf{I}}\cdot\hat{\mathbf{S}}
\end{equation}
where $\mathbf{B}_0$ is the applied magnetic field, $g$ and $A$ are the isotropic $g$-factor and hyperfine coupling, respectively, and $D$ is the axial ZFS parameter.  No evidence of a transverse anisotropy was observed for any of the three compounds, consistent with the three-fold rotational symmetry of the molecules. $D$ exhibits a systematic trend through the series, with \textbf{1} possessing easy-axis type anisotropy ($D < 0$) while \textbf{2} and \textbf{3} exhibit easy-plane type anisotropy ($D > 0$). By contrast, the hyperfine coupling is almost identical across the family  (Table~\ref{tab_para}) .

\begin{figure*}[t!]
\centering
\includegraphics[width=2\columnwidth]{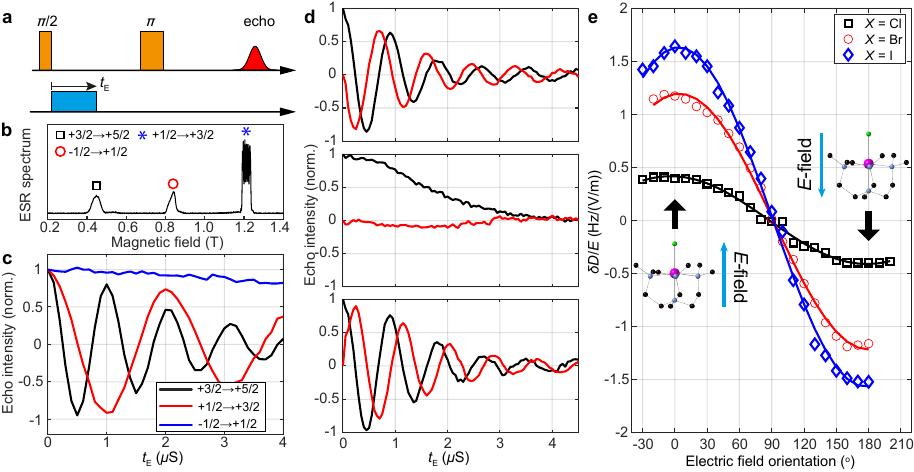}
\caption{Spin-electric coupling in Mn triangle molecules. \textbf{(a)} The microwave and \Efield pulse sequence measuring SEC in single crystals.  \textbf{(b)} The Q-band EDFS spectrum for \textbf{2} recorded at 3~K. \textbf{(c)} The in-phase spin echo signals for different $m_s$ transitions as a function of $t_\mathrm{E}$ recorded on \textbf{2}.  The data were recorded with both $B_0$ and the pulsed electric field parallel to the Mn-Br bond.  \textbf{(d)} The in-phase (black) and quadrature (red) echo signal for the +3/2 to +5/2 transition in MnBr with the electric field applied parallel (top),  perpendicular (mid) and anti-parallel (bottom) to the Mn-Br direction.  Note that the polarity of the quadrature signal is reversed for the top and bottom data,  consistent with a linear spin-electric coupling.  \textbf{(e)} Orientation dependence of the $E$-field-induced shift in the zero-field splitting parameter $D$ (errors are smaller than the symbol sizes). }
\label{fig_Efield}
\end{figure*}

We measured low-temperature spin relaxation times for \textbf{1} and \textbf{2} using magnetically diluted single crystals [Mn$_{0.001}$Zn$_{0.999}$(me$_6$tren)$X$]$Y_2$. The results are shown in Fig.~\ref{fig_relaxation} (see SI for single-crystal ESR spectra and corresponding energy diagrams). The spin lattice relaxation time $T_1$ for both molecules increases monotonically as the temperature falls, showing no sign of saturation down to our base temperature. At 3.5~K, $T_1$ for \textbf{1} (2.3~ms) is approximately 6 times that for \textbf{2} (0.36~ms). Such a difference in $T_1$ is likely due to the difference between the Mn-Cl (2.346~\AA) and Mn-Br (2.503~\AA) bond lengths: the longer Mn-Br distance leads to a weaker bond and lower energy vibrational modes, leading to faster spin-lattice relaxation rates at all temperatures that we studied. This is also consistent with the difference in the ESR spectra for \textbf{1} and \textbf{2}: while the hyperfine structure of \textbf{1} is clearly resolved for all transitions, we could only distinguish it for the $m_s = \pm1/2$ transition in \textbf{2} (Fig.~\ref{fig_Efield}\textbf{b}), indicating the presence of a significant $D$ strain (see SI). This also suggests that the structure of \textbf{2} might be more susceptible to an external \Efield,  potentially leading to a stronger SEC.

The phase coherence times ($T_m$) for \textbf{1} and \textbf{2} are described by a stretched exponential decay with a temperature dependent stretch parameter, indicating an interplay between multiple decoherence mechanisms~\cite{Wedge2012}. At 3.5~K both compounds show similar $T_m$ with their stretched parameters close to 2, suggesting that the decoherence is dominated by the nuclear spin bath surrounding Mn spins.  Upon raising the temperature, the coherence of \textbf{2} decreases rapidly, with the stretch parameter dropping below 1 at 5~K.  By comparison,  both $T_m$ and the stretch parameter for \textbf{1} remains almost temperature independent up to 10~K. 

The average Mn-Mn distance in our 0.1\% diluted crystals is less than 8~nm, so electron spin-spin dipolar interactions are of the order $\sim 1$~MHz). The difference in the temperature dependences of $T_m$ is likely related to the short $T_1$ associated with \textbf{2} (Fig.~\ref{fig_relaxation}\textbf{a}): $T_1$ relaxation in \textbf{2} leads to magnetic fluctuations in the local environment,  inducing contributions to the phase decoherence. Above 10~K, both the $T_m$ and the stretch parameter for \textbf{1} start to decrease rapidly with increasing temperature, suggesting that the nuclear spin bath is no longer the dominant decoherence source. 

\end{subsection}

\begin{subsection}{Spin-electric couplings}
We measured SECs for all three molecules by embedding a square dc \Efield pulse into a Hahn-echo sequence (Fig.~\ref{fig_Efield}\textbf{a}); we recorded the spin echo signal as a function of the duration/amplitude of the \Efield pulse~\cite{Liu2019,Liu2021} (more details in SI). Representative data (recorded on \textbf{2}) are shown in Fig.~\ref{fig_Efield}. The echo signals for the inter-Kramers transitions exhibit clear oscillations as a function of the duration of the \Efield pulse ($t_\mathrm{E}$), with the oscillation frequency for the $+5/2 \leftrightarrow +3/2$ signal almost exactly double that for the $+3/2 \leftrightarrow +1/2$ signal. By comparison, the $-1/2 \leftrightarrow +1/2$ transition shows only a weak SEC coupling. Measuring on different hyperfine peaks yields the same weak SEC. 

When the magnetic field $B_0$ is parallel to the magnetic anisotropy axis, the transition within the $\pm1/2$ doublet depends only on $g$ and $A$, whereas the inter-Kramers transitions also depend on the ZFS parameter $D$. Hence, the lack of $E$-field dependence for the $\pm1/2$ transition suggests both $A$ and $g$ show negligible SEC, and the oscillations observed with the inter-Kramers transitions are, therefore, atrtributable to the \Efield modulation of the anisotropy parameter $D$. This is further supported by the fact that the zero-field splitting for the $+5/2 \leftrightarrow +3/2$ transition, $4D$, is exactly twice of that for the $+3/2 \leftrightarrow +1/2$ transition, $2D$; therefore, an $E$-field induced modification in $D$, $\delta D$, lead to oscillation in the $+5/2 \rightarrow +3/2$ echo at frequency $4\delta D$, double that for the $+3/2 \rightarrow +1/2$ echo, $2\delta D$.

Both \textbf{1} and \textbf{2} crystallise in a polar space group with all molecules co-aligned. Consequently,  all molecules should exhibit the same linear response upon the application of an external \Efield, allowing us to measure both the amplitude and sign of \df. (The sign of \df is inaccessible for random orientated samples, e.g.\ frozen solutions~\cite{Liu2019} or single crystals with inversion related molecules~\cite{Liu2021}.) The in-phase and quadrature parts of the echo signal should follow $\cos{(\delta f_E t_\mathrm{E})}$ and $\sin{(\delta f_E t_\mathrm{E})}$, respectively, where the sign of \df is determined by the polarity of the quadrature component, as illustrated in Fig.~\ref{fig_Efield}\textbf{d}. When the orientation of the electric field is rotated from the Mn-Br (top) to the Br-Mn (bottom) direction, the quadrature part of the signal is inverted while the in-phase part remains virtually identical, as expected for a linear SEC.

The full orientation dependence of the SECs is mapped by rotating the \Efield against the crystals. We present the \Efield induced changes in $D$ for all three molecules for direct comparison (Fig.~\ref{fig_Efield}\textbf{e}). For all molecules the maximum SECs occur with the applied electric field parallel or anti-parallel to the Mn-$X$ bond, with a near-complete extinction of the effect for the perpendicular orientation. This highlights the importance of the molecular electric dipole: an \Efield is coupled to the molecular spin via its electric dipole. Hence, even though the triangular plane perpendicular to the Mn-$X$ bond also does not possess an inversion symmetry, allowing first order SEC~\cite{Trif2008,Trif2010,mims1976linear} by symmetry, an \Efield applied in this plane cannot couple to the spin efficiently owing to the lack of an electric dipole in this orientation. 

The observed SECs ($\sim 7$ Hz/(V/m) for \textbf{3}) are significant, especially considering that Mn(II) ions are typically associated with weak spin-orbit interaction due to their half-filled $3d^5$ outer shell. The coupling to the spin spectrum ($\delta f/E$) is much stronger than for molecular nanomagnets containing Mn(II) ($\sim 0.68$ HZ/(V/m))~\cite{Fang2022} and comparable with the SEC for a lanthanide-based molecule ($\sim 11$ Hz/(V/m)) with giant SOC~\cite{Liu2021}. Such strong SEC is likely due to the significant molecular electric dipole and the fact it is directly correlated to the molecular magnetic anisotropy. 

Despite the fact that $D < 0$ for $X =$ Cl and $D > 0$ for $X =$ Br and I, we note that $\delta D < 0$ for all three compounds when an electric field is applied pointing from the $X$ halogen ion towards Mn$^{2+}$. Such behaviour showcases the possibility of controlling magnetic anisotropy and SEC independently, allowing the design of molecular nanomagnets with strong SEC while maintaining operability within the microwave frequency range convenient for (quantum) information technologies. This can be understood qualitatively by considering the origin of the magnetic anisotropy and symmetry of their electronic states (see the section below).

\end{subsection}

\begin{subsection}{Electronic structure calculations}

We performed wavefunction-based \textit{ab initio} calculations to understand the origin of $D$ and its interaction with an external \Efield \cite{Maurice2009,Bouammali2021,Bouammali2021B,Pradines2022}. The ZFS parameters for all molecules (without external \Efield) are calculated using two geometries: the X-ray structures, and the molecular geometries optimised in density function theory (DFT) while preserving the $C_3$ symmetry. Both calculations reproduce the trend of $D$ observed in ESR measurements, i.e.\ the ZFS shifts from easy-axis type ($D < 0$) to easy-plane type ($D > 0$) as the halogen atom varies from Cl to I. Here we focus on results obtained using the DFT-optimised geometry as it allows us to investigate the \Efield induced distortions to the geometry of the molecules (see below). 

Detailed analysis was performed with \textbf{1} and \textbf{3} to rationalise the origin of the ZFS. For a high spin Mn(II) ($S = 5/2$) ground state, all five $d$ orbitals are singly occupied, leading to a sextuplet ground state $^6A$. Therefore, the ZFS can only emerge due to interactions between the electronic ground state and the excited quadruplet states, $^4Y^i$, via SOCs. It is worth noting that the spin-spin contribution to $D$~\cite{Duboc2007} which is considered in the calculations is very small. 

For an analysis purpose, we can consider the second order perturbation expression of the SOC contributions. The SOC interaction between the ground electronic state of $\ket{^6A_{m_{s}}}$ with the spin projection of $m_{s}$ and an excited electronic state of $^4Y^i$ in the $m_{sl}$ component leads to a contribution to $D$ of the ground state, $C(D)[^4Y^i_{m_{sl}}]$:
\begin{widetext}
\begin{equation}
\label{CD}
C(D)[^4Y^i_{m_{sl}}] =
\sum_{k}\frac{\bra{^6A_{m_{s}}}\zeta_k[(\hat{L}^+_k\hat{S}^-_k +  \hat{L}^-_k\hat{S}^+_k)/2 + \hat{L}^z_k\hat{S}^z_k]\ket{^4Y^i_{m_{sl}}}^2}{\mathcal{E}(^4Y^i)}
\end{equation}
\end{widetext}
where the sum runs over all electrons $k$ of the $d$ shell. $\ket{^4Y^i_{m_{sl}}}$ is the $^4Y^i$ excited state with the spin projection of $m_{sl}$. $\mathcal{E}(^4Y^i)$ is the energy of the $^4Y^i$ excited state with respect to the ground state and $\zeta_k$ is the SOC constant that depends on the two orbitals involved in the excitation. The sum of the contributions of all $^4Y^i$ excited states, $\sum{C(D)}$, leads to the ZFS.  \textit{Ab initio} calculations show that the main contributions to $D$ arise from the ten excited quadruplet states. Among them, four doubly degenerate states $E^i$ ($i = 1$ to 4) that couple to the ground state through the $(\hat{L}^+_k\hat{S}^-_k +  \hat{L}^-_k\hat{S}^+_k)/2$ term lead to negative contributions to $D$ and the two non-degenerate states $A^i$ ($i$ = 1 or 2) that couple to the ground state through $\hat{L}^z_k\hat{S}^z_k$ lead to positive contributions to $D$~\cite{Suaud2020}. 

\begin{table}[b]
\centering
\caption{Spin Hamiltonian parameters. The theoretical results for the zero-\Efield $D$ values are calculated using both the X-ray structure ($D_{\textrm{XR}}$) and the structure optimised using DFT ($D_{\textrm{DFT}}$). $\delta D/\delta E$ calculated using three different cases as described in the main text.}
\label{tab_para}
\renewcommand{\arraystretch}{1.5}
    \begin{tabular}{p{12em}|p{4em}|p{4em}|p{4em}}
    \hline
     Molecule & \textbf{1} & \textbf{2} & \textbf{3} \\
    \hline
   \multicolumn{4} {l}{Experiment}\\
   \hline
    $D$ (cm$^{-1}$) & -0.168 & +0.188 & +0.55 \\
    $A$ ($\times 10^{-3}$cm$^{-1}$) & 7.3 & 7.3 & 7.1 \\
    $\frac{\delta D}{dE}$ (Hz/(V/m)) & -0.42 & -1.20 & -1.70 \\
    \hline
    \hline
    \multicolumn{4} {l}{Theory, zero-\Efield $D$}\\
\hline
    $D_{\textrm{XR}}$ (cm$^{-1}$) & -0.172 & -0.057 & +0.098 \\
    $D_{\textrm{DFT}}$ (cm$^{-1}$) & -0.130 & +0.005 & +0.174 \\
    \hline
    \hline
    \multicolumn{4} {l}{Theory,  $\frac{\delta D}{dE}$ (Hz/(V/m))}\\
    \hline
    \textbf{(a)} electronic effect only& -0.043 & -0.112 & -0.239 \\
    \textbf{(b)} geometry effect only& -0.192 & -0.278 & -0.498 \\
    \textbf{(c)} both effects& -0.234 & -0.390 & -0.735 \\
    \hline
    \end{tabular}%
\end{table}%

The excitation energies in \textbf{1} and \textbf{3} are driven by the ligand field and follow the halogen spectrochemical series. However, while for many series of complexes, the excitation energies govern the magnitude and nature of $D$, the variation of the SOCs plays here the most important role. Indeed, one may notice that the increase or decrease of the contributions to $D$ of an excited state is directly correlated with the decrease or increase of the SOCs. The variation in SOCs can have two origins: either the coefficient on the $^4Y$ state determinants involved in the coupling varies between \textbf{1} and \textbf{3}, or the spin-orbit constants $\zeta_k$ vary. In the present case, both variations need to be considered. However, the dominant effect concerns the spin-orbit constants. Indeed, for an excitation involving orbital with a $z$ component (i.e. pointing towards the halogen), the constant $\zeta_k$ is weaker for the iodine-containing complex than for the chlorine-containing one due to the relativistic nephelauxetic effect, inducing weaker couplings and therefore lower negative contributions. Concerning the $A^2$ state, it is essentially carried by the two excitations from $d_{xy}$ to $d_{(x^2-y^2)}$ and vice versa and the weight on these two configurations is larger in \textbf{3} than in \textbf{1}, inducing a stronger coupling and therefore a larger positive contribution. To summarise, the negative contributions to $D$ due to the quadruplet $E^i$ states decrease from \textbf{1} to \textbf{3}, whereas the positive contributions brought by the $A^i$ states increase, resulting in an overall ZFS shifting from easy-axis to easy-plane type, as experimentally observed.

The application of an \Efield modifies both the electronic structure and the geometry of the molecules, thus changes $D$. To appreciate the spin-electric effect due to each contribution individually, we calculate $D$ using the following three cases: \textbf{(a)} an \Efield only modifies the electronic structure, with the molecular geometry unperturbed; \textbf{(b)} an \Efield distorts the geometry of the molecule, leading to a new structure (optimised using DFT in the presence of the \Efield) with which $D$ is calculated; \textbf{(c)} $D$ is calculated using the new geometry in the presence of an \Efield affecting the electronic structure. To reduce relative digital errors in the \textit{ab initio} calculation, a strong electric field is used ($\sim10^9$ V/m),  significantly larger than those applied in experiments ($\sim10^5$ V/m). Nevertheless, the calculations produce linear \Efield dependence of $D$ (Fig.~\ref{fig_theory}a), allowing us to draw a direct comparison between calculations and experiments.

\begin{figure}[t]
\centering
\includegraphics[width=\columnwidth]{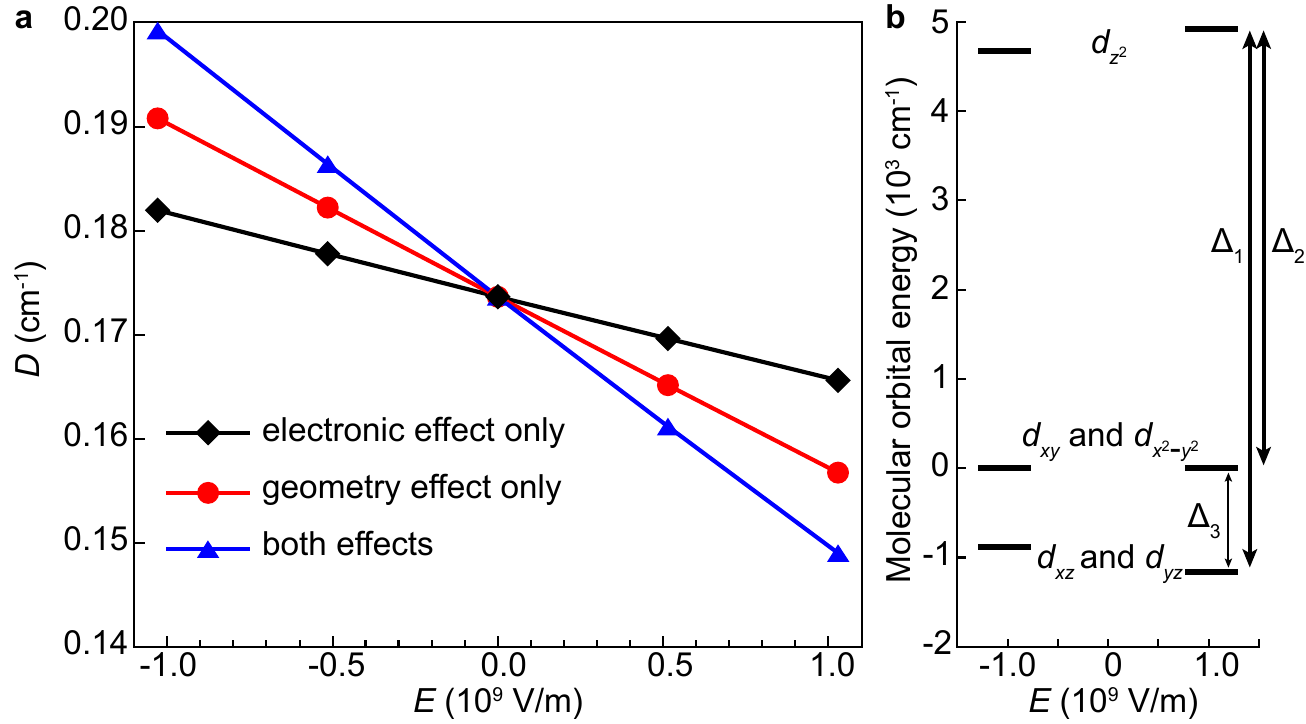}
\caption{\textbf{(a)} Theoretical calculation for \textbf{3} showing a linear SEC. A positive $E$ corresponds to an \Efield applied from I$^-$ to Mn$^{2+}$.  The calculations were performed with three configurations as described in the main text. \textbf{(b)} Molecular orbital energy diagram for \textbf{3} with the application of an \Efield.}
\label{fig_theory}
\end{figure}

Representative results for \textbf{3} are shown in Fig.~\ref{fig_theory}. When an \Efield is applied pointing from the halogen ion towards the Mn(II) ion, it distorts the molecular geometry such that the Mn--$X$ distance increases while the Mn--N bond length decreases. This changes the electronic structure of the molecule such that all energy differences between the orbitals increase, as shown in Fig.~\ref{fig_theory}\textbf{b}. Note that this increase is larger in the iodine-containing complex than in the chlorine one due to the larger polarisable character of iodine, i.e. a stronger deformation of the electronic cloud induces a larger geometric distortion of the molecule. More importantly, the application of an \Efield varies the SOCs, leading to a weaker $D > 0$ contribution by the $A_i$ states and a stronger $D < 0$ contribution by the $E_i$ states. These two modulations combine constructively and give rise to the overall \Efield-induced modulation of $D$. 

The results are summarised in Table~\ref{tab_para}. The calculations successfully reproduce the trend observed in experiments, with increasing effects when the halogen is changed from Cl to I. This can be understood by recognising that Cl$^-$ is less polarisable than I$^-$: applying an \Efield leads to a larger distortion in \textbf{3} and a stronger modulation of $D$, despite the SOC constant being stronger for \textbf{1}. Crucially, our analysis reveals that the distortions to the molecular geometry play the major role for all three molecules \cite{Liu2021}. Finally, we note that compared to the optimised structures used in the calculation, which are obtained considering single molecules in vacuum, the actual crystal structure contains counterions that can lead to larger distortions. Therefore, it is conceivable that the calculations underestimate the electric field effect. Nevertheless, the theoretical results are in reasonable agreement with the experimental data.
\end{subsection}
\end{section}

\begin{section}{Conclusions}
Our main findings are that it is possible to control the ZFS by electric field through and spin electric effect by prudent design of molecular complexes, and that we can generate a significant spin electric effect without involving a strong SOC. The theoretical analysis showcases the importance of geometry distortions in driving spin electric effect, and the possibility of harnessing the competing interactions to tune the ZFS and spin electric effect independently. 
Based on these findings we propose that lanthanide-containing complexes with an axial iodide ligand may present large SECs~\cite{Norel2018}. However, to be used as electrically addressable quantum bits, an ESR transition in the microwave energy domain is desirable, which is not straightforward for lanthanides. Our work suggests routes to QIP technologies based on molecular design principles, with electrically addressable molecular qubits compatible with current microwave techniques.
\end{section}

\bibliography{Mntriangleref}

\end{document}